\documentclass[a4paper,12pt]{article}      
\usepackage{amsmath,amssymb,amsfonts}      
\usepackage{graphicx}                      
\usepackage{color}                         
\usepackage[colorlinks = true,
linkcolor = red,
urlcolor  = blue,
citecolor = blue,
anchorcolor = black]{hyperref}\usepackage{eucal} 
\usepackage{appendix}
\usepackage{array} 
\usepackage{lscape}
\usepackage[english]{babel}
\usepackage[left=1.25in,right=1.25in,top=1.25in,bottom=1.25in]{geometry}
\usepackage{setspace}
\usepackage{parskip}
\usepackage{float}
\usepackage{longtable}
\usepackage{tikz}
\usepackage[T1]{fontenc}
\usepackage{rotating}
\usepackage{caption}
\usepackage{placeins}
\usepackage{placeins}
\usepackage{algorithmic}
\usepackage[ruled,vlined]{algorithm2e}
\usepackage{bm}
\usepackage{lineno}
\usepackage{booktabs}
\usepackage{natbib} 
\usepackage{verbatim}
\usepackage{pdflscape}
\usepackage{caption} 
\usepackage{authblk}
\usepackage{times}

\usetikzlibrary{arrows,positioning} 
\usetikzlibrary{shapes.geometric, arrows}
\title{Environmental policy in the context of complex systems: Statistical optimization and sensitivity analysis for ABMs}
\author[a,*]{Dylan Munson}
\author[b]{Arijit Dey}
\author[b]{Simon Mak}

\affil[a]{Sanford School of Public Policy, Duke University}
\affil[*]{Corresponding author: Dylan Munson, dylan.munson@duke.edu, 201 Science Drive, Durham, NC 27708}
\affil[b]{Department of Statistical Science, Duke University}
\date{}

\begin{document}

\begin{singlespace}
\maketitle
\end{singlespace}

\begin{singlespace}
\begin{abstract}
\noindent Coupled human-environment systems are increasingly being understood as complex adaptive systems (CAS), in which micro-level interactions between components lead to emergent behavior. Agent-based models (ABMs) hold great promise for environmental policy design by capturing such complex behavior, enabling a sophisticated understanding of potential interventions. One limitation, however, is that ABMs can be computationally costly to simulate, which hinders their use for policy optimization. To address this, we propose a new statistical framework that exploits machine learning techniques to accelerate policy optimization with costly ABMs. We first develop a statistical approach for sensitivity testing of the optimal policy, then leverage a reinforcement learning method for efficient policy optimization. We test this framework on the classic ``Sugarscape'' model, an ABM for resource harvesting. We show that our approach can quickly identify optimal and interpretable policies that improve upon baseline techniques, with insightful sensitivity and dynamic analyses that connect back to economic theory.\\

\vspace{0.5cm}

\end{abstract}
\end{singlespace}

\clearpage

\section{Introduction}\label{introduction}

Disorder is at least as old as, and, according to one of the most popular theories, more fundamental than time itself (\citealt{rovelliOrderTime2018}).  While the question of to what extent this conception of entropy applies directly to social systems rather than only through physical phenomena is still open (\citealt{mavrofidesEntropySocialSystems2011}), one notable trend in the social sciences has sought to embrace a certain form of disorder in its approach to this type of research, namely, ``complexity.''

``Complexity'' is a conceptually sticky term with no clear, single definition.  One common notion of complexity places it at the ``edge of chaos,'' where deterministic systems with strong sensitivity to initial conditions give rise to extremely complicated, hard to parse behavior (see, for example, the figure on page 10 of \cite{strogatzNonlinearDynamicsChaos2018}).  \cite{mitchellComplexityGuidedTour2009}, however, distinguishes several common properties of such systems and figures them as ``system[s] in which large networks of components with no central control and simple rules of operation give rise to complex collective behavior, sophisticated information processing, and adaptation via learning or evolution'' (p. 13).\footnote{Such a definition applies specifically to complex \textit{adaptive} systems, which are almost entirely the focus of this work.}  The three properties noted at the end of this definition seem broad enough to be almost redundant but, upon further reflection, ground the study of such systems in a very specific tradition of social scientific research, namely computational social science.

Computational social science (CSS) can be broadly thought of as a methodological variation on more traditional approaches to sociological and economic problems, applying computational principles instead of closed-form and equation-based ones.  These principles in particular become of use when dealing with \textit{complex data}--data which does not admit easy unpacking with closed-form equations because of feedbacks, nonlinearities, causal loops, psychological intricacies, etc.--data which is also usually ``large-scale'' and often simulated in nature (\citealt{lazerComputationalSocialScience2020}).  If this definition seems opaque, it is largely because the types of methods subsumed under the heading of ``computational social science'' are diverse and wide-ranging.  This paper will focus hereafter almost exclusively on one of these methods, agent-based modeling (ABM).

In particular, this attention to ABMs is warranted by increasing interest in their application to problems of public policy; we here focus on more specific questions of \textit{environmental} policy, a space within the ABM literature which has undergone greater exploration in recent years but still lacks full methodological clarity.  We here hope to contribute to this burgeoning literature by demonstrating both the practicability of applying ABMs to environmental policy questions, and by applying statistically rigorous techniques to the analysis of model outputs, thus solidifying the procedural underpinnings of this research space.

All that said, it would be a gross misrepresentation of prior research to say that no attempts have been made at applying the science of complex systems, in any form including those of agent-based models, to questions of governance and policy.  One of the foundational works of computational social science, Epstein and Axtell's \textit{Growing Artificial Societies}, notes the policy implications of the agent-based models of price dynamics presented in the work.  They cite in particular the power of their non-equilibrium results to challenge conventional wisdom about the operations of markets (\citealt{epsteinGrowingArtificialSocities1996}).  Indeed, it has been the argument of many researchers working on the application of complexity to social systems that it is precisely here where the power of such an approach lies: by allowing for adaptive and dynamic interactions between micro-level actors within the model framework, richer results and more precise policy statements may be made.  Existing work on the global sensitivity and uncertainty analysis of ABMs includes \cite{fonoberovaGlobalSensitivityUncertainty2013}; work on statistical inference of parameters has also been done (\citealt{g.v.bobashevUncertaintyInferenceAgentBased2010}).  We provide a more comprehensive literature review of ABMs for policymaking in the next section.

One key bottleneck with ABMs of complex systems, however, is that their simulation can be computationally costly. Such a cost arises from the need to carefully model for complex and fine-scale interactions in the system. Even for the classic ``Sugarscape'' model (\citealt{epsteinGrowingArtificialSocities1996}) investigated later in section \ref{results}, an ensemble of hundreds to thousands of simulations can take multiple hours to perform. The use of such costly ABMs for policy optimization can thus be computationally very intensive, especially when many policy levers are considered. This cost can hinder the promise of ABMs for policy optimization in practical applications. To address this, we propose a new statistical framework that leverages machine learning techniques to accelerate policy optimization with costly ABMs. In particular, our framework makes use of flexible learning models and reinforcement learning techniques that work well in this ``black-box'' setting (\citealt{frazier2018tutorialbayesianoptimization}), where data are obtained from costly simulator models. We will first develop a statistical approach for testing the sensitivity of the optimal policy with respect to state parameters, then make use of a Bayesian optimization reinforcement learning approach (\citealt{chen2024hierarchical}) for efficient policy optimization from the ABM. The effectiveness of this framework will be demonstrated on the classic ``Sugarscape'' model in \cite{epsteinGrowingArtificialSocities1996}, where we show that optimal and interpretable policies can be identified quickly, with insightful sensitivity and dynamic analyses that can be related back to economic theory.

We begin in section \ref{review} with a review of existing literature at the intersection of complex systems and policy analysis.  Section \ref{methods} presents the proposed statistical framework for sensitivity testing and policy optimization with ABMs.  Section \ref{model} then extends the Sugarscape model in \cite{epsteinGrowingArtificialSocities1996}, which we use as a proof-of-concept ABM for applying the proposed framework.  Section \ref{results} investigates the policies optimized using our method compared to baselines, along with corresponding sensitivity and dynamic analyses. Section \ref{conclusion} concludes.

\section{Background and Literature Review}\label{review}

It has long been understood that human-environment systems and interactions are ``complex'' in the colloquial sense of that word.  The extent to which this understanding has spilled over into a more precise, scientific one of complexity is less clear.  It is to interdisciplinary work, rather than work from a purely social or naturalistic view of the world, that we must turn to truly see this bridge materialize.

As just one introductory example, in his sweeping work \textit{The Great Transition: Climate, Disease and Society in the Late-Medieval World}, economic historian Bruce M.S. Campbell assesses the way in which a variety of social, economic, political, and environmental factors aligned to launch a fateful turn in the fortunes of Eurasia around the 14th century.  Not least amongst these were the repeated epidemics of the Black Death, long known to have been pivotal in shaping the course of European history.  Campbell, however, goes beyond a myopic analysis of these events, including as well solar cycle data from sunspot observations and temperature anomalies extracted from bubbles trapped in ice, to name just two.  The multiple data sources appealed to in the work, while indicative of the need for interdisciplinary understanding, is not the point to emphasize here, however; it is the way in which Campbell highlights the \textit{complexity} of the interactions between different factors, such as disease and climatic anomalies, as being paramount in determining the course of his ``Great Transition.''  No single event or effect was responsible for the change in Europe's fortunes, but it was instead the interactions between them that led to system-level changes, which in turn shaped the course of civilization on the continent.  As the author puts it, ``[a]t any given juncture several different outcomes were possible depending upon the precise configuration of human and environmental forces'' (\citealt{campbellGreatTransitionClimate2016}, p. 3).  This notion is startlingly close to that of \textit{sensitive dependence on initial conditions} which is so vital to an understanding of scientific chaos and, by extension, complexity.

Can interactions between people and their environment, such as those that brought about the medieval Great Transition, be more generally considered complex, then?  We argue, as others such as \cite{levinSocialecologicalSystemsComplex2013}, \cite{annurev:/content/journals/10.1146/annurev-environ-110615-085349}, and \cite{preiserSocialecologicalSystemsComplex2018} have before, that the answer is yes.  In the first of these works, the authors posit that social-ecological systems emerge from collective behavior which percolates to higher levels of the systems.  This thinking fits squarely in the tradition upon which ABMs and complex adaptive systems (CAS, complex systems in which system-level changes significantly occur through learning or evolution (\citealt{mitchellComplexityGuidedTour2009})) are built: micro-level interactions coalesce into processes which exhibit behavior that could not easily be predicted from an analysis of the individual agents themselves.  From a policy standpoint, effective management of such a system requires balancing redundancy, heterogeneity, and modularity to ensure the system is resilient to shocks and losses.

The authors further emphasize that nonlinearities present in such systems may give rise to tipping points not well predicted by non-CAS models, as in the case of reef systems, an argument also taken up by \cite{annurev:/content/journals/10.1146/annurev-environ-110615-085349}.  Tipping points and regime changes are well within the purview of complexity and chaos embedded, as they are, in the study of bifurcations (see, for example, chapter 3 in \cite{strogatzNonlinearDynamicsChaos2018}).  They are also a topic of major concern in climate change modeling, with small perturbations to various environmental systems due to human activity potentially leading to qualitative changes in the behaviors of those systems (\citealt{lentonTippingElementsEarths2008}).  Although climate modeling is not the focus of the present study, the fact remains that understanding coupled human-environment systems requires an acceptance of the fact that underlying dynamics are likely nonlinear and complex, and thus difficult to model, especially with more traditional approaches.

The benefit of a computational approach (note that numerically solved non-linear models have long been a norm in global climate modeling (\citealt{mcguffieFortyYearsNumerical2001})) is that complex (or in the case when closed form systems are available, non-linear) processes may be modeled, as long as computational power is sufficient.  Indeed, as computing capacity has increased vastly in recent decades, more disciplines have begun to recognize the potential of complexity in solving pressing methodological questions.  The spheres of public policy, including environmental policy, have been among these.

During a 2001 symposium on the topic of the use of ABMs for the simulation of public policy, \cite{lempertAgentbasedModelingOrganizational2002} argued for the use of such models, coupled with ``new analytic approaches for decision-making under conditions of deep uncertainty,'' to simulate phenomena which cannot be represented with simple mathematical structures.  In fact, it is exactly these types of advances in the methodology and systematization of the use of ABMs that will allow their use to tackle rich problems of environmental concern; it is also the type of contribution which this paper hopes to make.  More recent proposals have called for the use of ABMs to shape food policy (\citealt{giabbanelliUsingAgentBasedModels2017}) and transportation governance (\citealt{maggiUnderstandingUrbanMobility2016}).  Still, empirical applications have been lagging.

One area in which some progress has been made on this front is in the sphere of environmental, and more particularly agricultural, policy.  A commonly used model in this literature is the Agricultural Policy Simulator (AgriPoliS) (\citealt{happeAgriculturalPolicySimulator2004}).  The model, designed with a German context in mind, is meant to overcome the limitations of overly-aggregated macroeconomic models which make the analysis of policies impacting the micro-level difficult.  The model, however, does not necessarily have policy optimization itself in mind; instead, farmers (who are the agents in the model) by assumption know about major policy changes \textit{before} they take effect.  This makes flexible analysis of exogenous large-scale policy shifts (e.g. regime changes) through sensitivity analysis relatively straightforward (see, for example, \cite{happeAgentbasedAnalysisAgricultural2006}), but the assumptions built in to the model would presumably not allow internal changes in policy to be optimized.  Similarly, the Regional Multi-Agent Simulated (RegMAS) simulates local \textit{responses} to policy changes, making it a valuable tool for scenario analysis but not necessarily for policy choices (\citealt{lobiancoRegionalMultiAgentSimulator2010}).  Policies are in fact read into this model from external sources rather than being levers operated within the model machinery itself, as in the model we present below.  On the other hand, sensitivity analysis to various innovations and changes in model parameters can allow some form of rudimentary policy tuning; steps in this direction seem to be made by the designers of the Mathematical Programming-based Multi Agent Systems (MP-MAS) (\citealt{schreinemachersAgentbasedSimulationModel2011}).  That model has been used, for example, to measure the effects of better access to credit and improved technologies for farmers; the model is in fact able to quantify the effects of such policy changes on various model outputs such as the poverty rate (\citealt{schreinemachersSimulatingSoilFertility2007}).  As can be seen in this brief survey, scenario analysis is a common theme with agent-based models, and does allow for policy tuning, but true policy optimization has not been extensively tested in the literature to our knowledge. 

It is also worth noting that other agent-based models have specifically taken up the issue of adaptation to climate change, including amongst agriculturalists.  While a thorough review of this literature will thus be left for that work, it has been argued that MAS systems of the type previously mentioned, as well as ABMs, are well-suited for making policy decisions because of their ability to capture uncertainty and adaptation within the framework of the model itself (\citealt{bergerAgentbasedModellingClimate2014}).  Such simulators have been used to explore, for example, shifts in adaptive capacity due to climate change in Ethiopia (\citealt{hailegiorgisAgentBasedModelRural2018}).

Despite such advances, further methodological advances are needed to fully realize the promise of ABMs for policy crafting.  On the one hand, it is worth noting from the outset that policy should not be designed exclusively based on the recommendation of a single ABM or simulation ensemble (indeed, policy advice should almost never be tied to a single source in any case).  ABMs and a complex systems approach have the ability to challenge conventional notions of economic and environmental dynamics, but should be coupled with other approaches to craft intelligent, fair, and well-founded policy.  On the other hand, such methods provide a unique opportunity to simulate a \textit{continuum} of policy options and optimize over these possibilities.  Figure \ref{schematic} displays one possible way of understanding the relationship between theoretical and empirical foundations and the use of ABMs for the design and implementation of policies.  With the micro-founded fundamentals of such models in hand, simulation results can provide rich analysis of the impacts of shifting a given policy lever (or set of levers) on macro, distributional, and agent-level outcomes.  Specifically, the methods that we develop in this paper can be used to (1) test for sensitivity of model outputs to various potential policy-moderating state variables, and then (2) if policy outcomes are determined to be sensitive to such variables, more thoroughly optimize over policy options using carefully-implemented machine learning techniques.

To achieve this policy optimization goal with ABMs, one key challenge is the computational cost of ABM simulations: each run can require minutes or even hours to perform for sophisticated models. This can hinder the optimization of such models for policy crafting, as it limits the number of policy choices that one can test from the ABM. To address this, we propose in the following a new framework for policy optimization, which incorporates machine learning techniques to accelerate this optimization procedure.  Our work contributes to the kind of policy analysis workflow described above, by leveraging statistical sensitivity testing and reinforcement learning techniques that work well with limited simulation runs from the ABM.  Our methodology is highly adaptable to a multitude of use cases beyond the highly stylized environmental and natural resource model we use in this work.  We thus contribute not only to the literature on agent-based modeling and statistical optimization of such models, but also the field of policy analysis more generally, where such models can be of great use (as is evidenced by the literature above).

\section{Methods: A Statistical Framework for Sensitivity Testing and Optimization of ABMs}\label{methods}

We here present the proposed ML-based statistical framework for sensitivity testing and subsequent policy optimization using complex ABMs. We first outline in subsection \ref{sensitivity_analysis} our sensitivity testing approach, which leverages statistical hypothesis testing techniques with a flexible ML model called a Gaussian process \citealt{gramacySurrogatesGaussianProcess2020}. We then describe in subsection \ref{bo} a reinforcement learning approach that uses this ML model to guide policy optimization from the costly ABM.

We first establish some notation. Suppose the ABM takes in two types of input variables: (i) policy (or ``choice'') variables $x \in \mathbb{R}^d$, which parametrize a discrete or continuous policy space $\mathcal{P}$, and (ii) state variables $\mathbf{\theta} \in \mathbb{R}^p$, which parametrize what we will sometimes refer to as system ``state variables,'' roughly in line with the corresponding concept from optimal control theory (\citealt{leonardMaximumPrinciple1992}). For example, to investigate agricultural resilience, an ABM may take in inputs $x$ that dictate different adaptation subsidy policies, as well as inputs $\theta$ that model for varying household and environmental characteristics.

Given variables $x$ and $\theta$, we model the simulated response from the ABM as a realization from the random variable $G(x,\theta)$, as such a response is inherently stochastic due to randomness in initial model states, bounded rationality in behavioral patterns, and other randomness inherent in agent decisions and the model environment. For example, $G(x.\theta)$ may model the distribution of wealth or welfare in a population, from which the ABM output is simulated. A reasonable formulation of the policy optimization problem might then be:
\begin{equation}
x^*(\theta) := \arg\min_x f(x,\theta), \quad f(x,\theta) := \Psi\{G(x,\theta)\},
\label{eq_general_stat}
\end{equation}
where $f$ is the target objective function to minimize. Here, $\Psi$ is a functional that maps the random variable $G$ onto a real number. Common choices of $\Psi$ include the distribution mean $\Psi(G) = \mathbb{E}(G)$ (e.g., the average wealth in the population), or the distribution 5\%-quantile $\Psi(G) = Q_{5\%}(G)$ (e.g., the wealth level below which 5\% of the population falls). In our later experiments, we employ the distribution mean for $\Psi$ as a proof-of-concept. Our proposed approach can be used analogously for either minimizing or maximizing $f$; for brevity, we presume minimization in what follows.

A key challenge for the policy optimization problem \eqref{eq_general_stat} is that, for a complex ABM, each evaluation of the objective function $f(x,\theta)$ can be computationally costly. Given a choice of $(x,\theta)$, the evaluation of $f(x,\theta)$ requires many simulations from the ABM, i.e., sample draws from the distribution $G(x,\theta)$, to reliably compute the functional $\Psi\{G(x,\theta)\}$; we provide guidance on the number of sample draws later in section \ref{sec:algorithmic_statement}. For sophisticated ABMs, the cost of performing each simulation run can be non-negligible, particularly with many agents in the model, fine-grained agent representations (\citealt{schwarzAgentbasedModelingDiffusion2009}), or increased fidelity in modeling physical environments (e.g., water flow dynamics; \citealp{dawsonAgentbasedModelRiskbased2011}). This means a single evaluation of $f(x,\theta)$ can require minutes or even hours, which greatly limits the number of $(x,\theta)$ combinations (i.e., different combinations of policies and state parameters) that one can test from the ABM. This makes the target problem of sensitivity testing for $f$ and its subsequent policy optimization highly difficult problems.

In the ML literature, the property that $f$ is unknown and costly to evaluate is referred to as a ``black-box'' function (\citealt{frazier2018tutorialbayesianoptimization}). The analysis and optimization of such black-box functions is challenging, and require carefully-designed ML techniques that work well with limited evaluations of $f$. To this end, we present two ML-based approaches for effective sensitivity testing and policy optimization from costly ABMs.

\subsection{Sensitivity Testing}\label{sensitivity_analysis}

First, we consider a problem inherent to the black-box nature of the function $f$: is the optimal choice of $x$, $x^{\star}$, dependent on, or \textit{sensitive to} a particular value (or values of) the state parameter(s) $\theta$?  In other words, if $\theta$ may take on certain values, or perhaps in a formulation that is more relevant to the questions at hand, if the actual value of $\theta$ is uncertain but known to be within some range, is $x^{\star}(\theta) \approx C$ for all choices of $\theta$, where $C$ is some constant? Or is $x^{\star}$ truly a function of $\theta$? If the answer to this last question is ``yes,'' then optimal policy choices could vary drastically over even minor changes to the model state and initial conditions.  For this reason, we would need to be exceptionally careful in optimizing over ensembles of model runs which may have different initial conditions, and over different possible values of $\theta$.

The above notion of sensitivity is directly related to the idea of additivity for $f$. The function $f(x,\theta)$ is \textit{additive} if it can be written as $f(x,\theta ) = g_1(x) + g_2(\theta)$ for some functions $g_1$ and $g_2$. Note that the additivity of $f$ immediately implies that the optimizer of $f$, namely $x^{\star}$, is not dependent on $\theta$.  However, if the function is instead \textit{non-additive}, i.e., $f(x,\theta ) = g_1(x) + g_2(\theta) + g_3(x,\theta )$ for some appropriate functions $g_1, g_2, g_3$, where $g_3(x,\theta)$ is non-zero, then the optimizer $x^{\star}(\theta)$ may indeed be dependent on $\theta$.  We present next a statistical procedure for testing additivity on the black-box function $f$, which allows us to investigate the sensitivity of the optimal solution $x^{\star}$ to the state variables $\theta$.

Suppose we simulate data of the form $\mathcal{D} = \{f(x_1, \theta_1), \cdots, f(x_n, \theta_n)\}$ from the ABM, where $(x_1, \theta_1), \cdots, (x_n, \theta_n)$ are tested combinations of policy and state variables. Here, the number of tested combinations $n$ is again limited due to the costly nature of function evaluations. Our statistical hypothesis test makes use of Gaussian process (\citealt{gramacySurrogatesGaussianProcess2020}) models, which are flexible Bayesian ML models broadly used in scientific applications, e.g., aerospace engineering (\citealt{miller2024expected}) and space science (\citealt{li2025prospar}). A key appeal of GPs is that they provide effective, reliable and uncertainty-aware learning with a limited sample size $n$, all of which are needed for our sensitivity testing and policy optimization goals here.

Formally, a GP model on $f$ can be denoted as:
\begin{equation}
f \sim \text{GP}\{\mu,K(\cdot,\cdot)\},
\label{eq:gp}
\end{equation}
where $\mu$ is a scalar mean hyperparameter, and $K(\cdot,\cdot)$ is a scalar-valued kernel function. A popular kernel choice is the squared-exponential kernel (\citealt{gramacySurrogatesGaussianProcess2020}):
\begin{equation}
K((x,\theta),(x',\theta')) = \sigma^2 \exp\left\{ - \sum_{l=1}^d \left(\frac{x_{[l]} - x_{[l]}'}{\psi_{x,l}}\right)^2 - \sum_{l=1}^p \left(\frac{\theta_{[l]} - \theta_{[l]}'}{\psi_{\theta,l}}\right)^2 \right\},
\label{eq:sqexp}
\end{equation}
where $x = (x_{[1]}, \cdots, x_{[d]})$ and $\theta = (\theta_{[1]}, \cdots, \theta_{[p]})$ are the vectors of policy and state variables, respectively. Here, $\sigma^2$ is a kernel variance hyperparameter, and $\{\psi_{x,l}\}_{l=1}^d$ and $\{\psi_{\theta,l}\}_{l=1}^p$ are kernel length-scales that model the importance of different policy and state variables on $f$, respectively. The model hyperparameters $\mu$, $\sigma^2$, $\{\psi_{x,l}\}_{l=1}^d$ and $\{\psi_{\theta,l}\}_{l=1}^p$ can be learned from the simulated data $\mathcal{D}$. The class of functions modeled by a GP of this form can be shown to be nonparametric and highly flexible (\citealt{van2011information}).

With this, suppose the true function $f(x,\theta)$ is additive in nature. Then a reasonable learning model for $f$ may be $f(x,\theta) = g_1(x) + g_2(\theta)$, where $g_1 \sim \text{GP}\{\mu_1,K_1(\cdot,\cdot)\}$ and $g_2 \sim \text{GP}\{\mu_2,K_2(\cdot,\cdot)\}$. Here, $K_1(x,x')$ is a kernel depending on only policy variables $x$, and $K_2(\theta,\theta')$ is a kernel depending on only state variables $\theta$. For kernel $K_1$ (or kernel $K_2$), the squared-exponential form \eqref{eq:sqexp} can be used with only the length-scale hyperparameters $\{\psi_{x,l}\}_{l=1}^d$ (or only the length-scale hyperparameters $\{\psi_{\theta,l}\}_{l=1}^p$).  We shall call this the \textit{null} model. Conversely, suppose the true $f$ is non-additive. Then an alternate model might be $f(x,\theta ) = g_1(x) + g_2(\theta) + g_3(x,\theta )$, where $g_1$ and $g_2$ follow the above GP models, and $g_3 \sim \text{GP}\{0,K_3(\cdot,\cdot)\}$, where the kernel $K_3((x,\theta),(x',\theta'))$ is over both $x$ and $\theta$. 
For kernel $K_3$, we adopt the squared-exponential form given in \eqref{eq:sqexp}, where its length-scale hyperparameters are distinct from those for $K_1$ and $K_2$.
We shall call this the \textit{alternative} model. All kernels $K_1$, $K_2$ and $K_3$ share the same kernel variance $\sigma^2$. Hyperparameters for both the null and alternative models can be learned from data by maximizing their likelihood functions (\citealt{casella2024statistical}).

Our hypothesis test for additivity can then be formalized as follows. Define the ``null hypothesis'' as the hypothesis where $f$ follows the above additive null model. Define the ``alternative hypothesis'' as the hypothesis where $f$ follows the broader non-additive alternative model. We adopt a \textit{likelihood ratio test} (LRT; \citealp{casella2024statistical}) to test whether there is sufficient evidence from data $\mathcal{D}$ to reject the null hypothesis of an additive $f$. The LRT uses the test statistic:
\begin{equation}
    \Lambda = -2 (l_0 - l_1),
    \label{eq:lrt}
\end{equation}
where $l_1$ is the log-likelihood of the alternative model fitted using data $\mathcal{D}$, with model hyperparameters learned via maximum likelihood, and $l_0$ is the log-likelihood of the null model fitted using data $\mathcal{D}$, with hyperparameters learned in a similar fashion. The full likelihood expressions under a GP can be found in \cite{gramacySurrogatesGaussianProcess2020}; they are omitted here for brevity. Note that a larger test statistic $\Lambda$ gives stronger evidence that additivity is violated for $f$, as this implies the log-likelihood under the alternative model is considerably larger than that under the null model. 

One can prove (\citealt{casella2024statistical}) that, under the null model, the test statistic $\Lambda$ follows approximately a chi-squared distribution $\chi^2_\nu$ with $\nu = d + p$ degrees-of-freedom, since the non-additive alternative model has $d + p$ additional hyperparameters over the additive null model. The p-value for this test can then be computed as the probability $\mathbb{P}(\chi^2_\nu > \Lambda)$. A small p-value suggests that, under the null hypothesis of an additive $f$, the probability of observing the data $\mathcal{D}$ is low. This thus provides statistical evidence against the additivity of $f$, and suggests that the optimal solution $x^\star$ may indeed be sensitive to the state variables $\theta$. Conversely, a large p-value gives little statistical evidence against the additive nature of $f$, which in turn suggests the optimal solution $x^*$ may be insensitive to $\theta$.

\subsection{Bayesian Optimization}\label{bo}

Suppose we find, from the above sensitivity test, that the optimal policy decision $x^\star(\theta)$ is indeed sensitive to the state parameters $\theta$. To further investigate this, we may wish to find, for different choices of $\theta$, how its corresponding optimal solution $x^*(\theta)$ from \eqref{eq_general_stat} changes. Again, finding $x^*(\theta)$ is a highly challenging problem, since $f$ is a black-box function with no closed-form expression and is costly to evaluate, requiring simulations from the ABM. To solve \eqref{eq_general_stat}, we need to leverage black-box optimization algorithms that work well with limited evaluations on $f$.

Bayesian optimization (BO; \citealp{frazier2018tutorialbayesianoptimization}) provides an effective solution. BO is a reinforcement learning approach for optimizing black-box functions, with successful applications in machine learning (\citealt{snoek2012practical}) and engineering design (\citealt{kim2025efficient}). For a fixed choice of $\theta$, the idea is to use existing evaluations on $f$ to construct a so-called ``acquisition function'' $a(x)$, which leverages a fitted learning model to quantify the attractiveness of a potential evaluation point $(x,\theta)$ for optimization. Next, the function $f$ is evaluated at the point that maximizes this acquisition function $a(x)$. The learning model is then refit with the new data from the ABM, and the above sequential sampling procedure is repeated until a satisfactory optimization solution is obtained.

We will employ a popular BO approach called Expected Improvement (EI; \citealp{jones1998efficient}) for optimizing $x^*(\theta)$, which leverages the GP model introduced earlier on the objective function $f(\cdot,\theta)$. Let $\theta$ be the fixed set of considered state variables. Suppose $f$ has already been evaluated at the input combinations\footnote{Here, the notation $x_i'$ is used to distinguish these evaluation points from those used in section \ref{sensitivity_analysis}.} $(x_1',\theta), \cdots,(x_m',\theta)$ from the ABM. The EI acquisition function is then defined as:
\begin{equation}
\text{EI}(x) = \Phi \left( \frac{f_{\rm min} - \hat{f}(x)}{s(x)} \right) (f_{\rm min} - \hat{f}(x)) + \phi \left( \frac{f_{\rm min} - \hat{f}(x)}{s(x)} \right) s(x).
\label{eq:ei}
\end{equation}
where $\Phi$ and $\phi$ are the cumulative and probability density functions of a standard normal distribution, respectively, and $\displaystyle f_{\rm min} = \min_{i=1,\cdots,m} f(x_i',\theta)$ is the smallest observed value of the objective function. Here, $\hat{f}(x)$ and $s(x)$ are the predicted value of $f(x,\theta)$ from the fitted GP model and its corresponding predictive standard deviation; see \cite{gramacySurrogatesGaussianProcess2020} for their full expressions. $\text{EI}(x)$ can further be viewed as the expected objective improvement from a new function evaluation at inputs $(x,\theta)$; see \cite{chen2024hierarchical}.


With this, the next evaluation point $(x_{m+1}',\theta)$ is selected to maximize the expected improvement acquisition, i.e., $x_{m+1}' = \arg\max_x \text{EI}(x)$. Points selected by maximizing $\text{EI}(x)$ capture the ``exploration-exploitation trade-off'' (\citealt{liu2025quip}) fundamental in reinforcement learning: the maximization of the first term in \eqref{eq:ei} targets $x$ with small predicted objectives $\hat{f}(x)$ (i.e., exploitation), whereas the maximization of the second term targets $x$ with large predictive uncertainties $s(x)$ (i.e., exploration). The ABM is then simulated at the optimized point $(x_{m+1}',\theta)$, the GP model is re-fit, and the above sequential sampling procedure is repeated until a good solution is found for optimizing \eqref{eq_general_stat}. For optimizing black-box ABMs, this BO procedure can be considerably more efficient computationally than random sampling of the policy variables, as we show later in section \ref{bo_results}.

\subsection{Algorithm Statements}
\label{sec:algorithmic_statement}

Algorithms \ref{alg:sensitivity} and \ref{alg:bo} outline the detailed steps for our proposed sensitivity testing and policy optimization approaches. We provide a brief discussion of each in the following.

Our sensitivity testing algorithm (Algorithm \ref{alg:sensitivity}) begins with selecting the initial evaluation points $\{(x_i,\theta_i)\}_{i=1}^n$ for the black-box function $f$. Following the rule-of-thumb for GP modeling (\citealt{loeppky2009choosing}), we recommend the number of evaluation points be set as $n = 10(d+p)$, where $d+p$ is the number of policy and state variables in the ABM. These points should further be selected from a Latin hypercube design (LHD; \citealp{stein1987large}), which offers improved performance compared to random sampling. The ABM is then run at each of these points, and the objective $f$ is evaluated via Equation \eqref{eq_general_stat}. Note that, for a given $(x,\theta)$, evaluating $f(x,\theta)$ requires many ABM simulation runs, i.e., sample draws from the distribution $G(x,\theta)$. In our experiments later, we find that 150 sample draws provide sufficient precision for good sensitivity testing and policy optimization performance. 

Next, using the simulated data from the ABM, namely $\mathcal{D} = \{f(x_i,\theta_i)\}_{i=1}^n$, we fit the GP models under the null (additive) and alternative (non-additive) hypotheses. Hyperparameters for these models are trained via maximum likelihood (\citealt{casella2024statistical}), where likelihood optimization is performed via the Adam optimizer (\citealt{kingma2014adam}). Model training of all GPs is performed via the \texttt{GPyTorch} package (\citealt{gardner2018gpytorch}) in Python. The LRT statistic $\Lambda$ is then computed from \eqref{eq:lrt}, and the test p-value is evaluated as the probability $\mathbb{P}(\chi^2_\nu > \Lambda)$. Finally, this probability is compared with a pre-set significance level $\alpha$ (typically taken to be 0.05) to determine whether there is enough statistical evidence to reject additivity in $f$.

Our Bayesian optimization algorithm (Algorithm \ref{alg:bo}) then aims to identify the optimal policy decision $x^*(\theta)$ for a fixed choice of state variables $\theta$. For initial model learning, we first evaluate $f$ at the points $\{(x_i',\theta)\}_{i=1}^{m}$, where the points $x_1', \cdots, x_m'$ are selected from an LHD. Here, a small number of initial evaluations $m$ is sufficient (e.g., $m=5$), as the Bayesian optimization procedure adds in sequential evaluations after. With this initial data, we fit a GP model on the objective function $f(\cdot,\theta)$, and optimize its hyperparameters via maximum likelihood (\citealt{casella2024statistical}). Next, we select the next evaluation point $x_{m+1}'$ by maximizing the acquisition function $\text{EI}(x)$ in \eqref{eq:ei} using the Adam optimizer (\citealt{kingma2014adam}), and evaluate $f$ at the new point $(x_{m+1}',\theta)$ via the ABM. The GP model is then re-fit with this new data, and the above sequential sampling procedure is repeated until a good optimization solution is found. For our later experiments, we find that $M=100$ iterations of this sequential procedure is sufficient for optimization, although this should be determined on a case-by-case basis given computational time constraints. In our implementation, Algorithm \ref{alg:bo} is performed using the Python package \texttt{BoTorch} (\citealt{balandat2020botorch}).

\begin{algorithm}[!t]
\caption{Sensitivity Testing of Optimal Policy}
\label{alg:sensitivity}
\begin{algorithmic}[1]
\REQUIRE Number of evaluation points $n$, significance level $\alpha$

\STATE Select evaluation points $\{(x_i,\theta_i)\}_{i=1}^n$ from an LHD.

\STATE Run the ABM at these evaluation points to obtain data $\mathcal{D} = \{f(x_i,\theta_i)\}_{i=1}^n$.

\STATE \textit{Null Model}: Using data $\mathcal{D}$, optimize hyperparameters for the additive GP via maximum likelihood using the Adam optimizer (\citealt{kingma2014adam}). Define $l_0$ as the log-likelihood of this model with optimized hyperparameters.

\STATE \textit{Alternative Model}: Using data $\mathcal{D}$, optimize hyperparameters for the non-additive GP via maximum likelihood using the Adam optimizer (\citealt{kingma2014adam}). Define $l_1$ as the log-likelihood of this model with optimized hyperparameters.

\STATE Compute the LRT statistic $\Lambda = -2 (l_0 - l_1)$, and compute the p-value $\varsigma = \mathbb{P}(\chi^2_\nu > \Lambda)$, where $chi^2_\nu$ is the chi-squared distribution with $\nu = d + p$ degrees-of-freedom.

\STATE If $\varsigma < \alpha$, there is statistical evidence that the optimal policy $x^*(\theta)$ is sensitive to state variables $\theta$. Otherwise, there is insufficient statistical evidence for sensitivity.

\end{algorithmic}
\end{algorithm}

\begin{algorithm}[!t]
\caption{Policy Optimization via Bayesian Optimization}
\label{alg:bo}
\begin{algorithmic}[1]
\REQUIRE Fixed state variables $\theta$, number of initial evaluation points $m$, number of sequential BO iterations $M$

\STATE Select initial evaluation points $\{(x_i',\theta)\}_{i=1}^{m}$ from an LHD.

\STATE Run the ABM at these initial points to obtain data $\mathcal{D}' = \{f(x_i',\theta)\}_{i=1}^{m}$.

\FOR{$t = 1, \cdots, M$}
    \STATE Fit a GP model on the data $\mathcal{D}'$, with model hyperparameters optimized via maximum likelihood.
    \STATE Optimize the next policy variables to be evaluated as $x_{m+t}' = \arg\max_x \text{EI}(x)$ using the Adam optimizer (\citealt{kingma2014adam}).
    \STATE Run the ABM at the new input point $(x_{m+t}',\theta)$ and evaluate $f(x_{m+t}',\theta)$.
    \STATE Update the training data $\mathcal{D}' \leftarrow \mathcal{D}' \cup \{f(x_{m+t}',\theta)\}$.
\ENDFOR

\STATE \textbf{Return:} The best observed policy $\displaystyle x^* = \underset{x \in \{x_1', \cdots, x_{m+M}'\}}{\arg\min} f(x,\theta)$.

\end{algorithmic}
\end{algorithm}

\section{Proof-of-Concept Model: Pollution on the Sugarscape}\label{model}

\subsection{Model Description}

We now apply the proposed framework for sensitivity testing and policy optimization to a proof-of-concept application. In what follows, we will adopt and extend the Sugarscape model, described in Chapter 4 of \cite{epsteinGrowingArtificialSocities1996}. The classic software for building such an agent-based model is NetLogo, but we have here used an extended version of the model constructed using the Python-based \texttt{mesa} framework (\citealt{python-mesa-2020}) to improve reproducibility. A version of the base \texttt{mesa} model, to which we made adjustments in order to incorporate pollution and mitigating policy options, is available at \url{https://mesa.readthedocs.io/latest/examples/advanced/sugarscape_g1mt.html}.

In the Sugarscape model, space is composed of a grid cell lattice on which agents move.  Each grid cell is endowed with a resource, called by convention ``sugar'' and ``spice,'' which are harvested by the agents and which regrow at a constant rate over time.  In our version of the model, the resources grow back at a rate of 1 unit per time step, up to a pre-specified maximum amount that differs between cells.  Agents will trade sugar for spice under certain conditions, and also move to new cells based on welfare-maximizing principles.\footnote{In particular: 1) agents trade when they encounter one another on the lattice, and their marginal rates of substitution (MRS) between sugar and spice are not equal; 2) agents ``search'' neighboring cells (within a specified horizon) and move to the one that will maximize their welfare.}  Agents harvest all the resources on a grid cell which they occupy at a given time step of the model.  They also must consume sugar and spice, and if one of these resources is exhausted for the agent, they die.  Thus the rule for agent movement, as given in \cite{epsteinGrowingArtificialSocities1996}, pp. 98-99, is:

\begin{enumerate}
	\item Look as far as vision permits in each cardinal direction.
	\item Taking as feasible only unoccupied positions in the lattice, find the nearest position that maximizes welfare.
	\item Move to that position.
	\item Harvest all sugar and spice at that location.
\end{enumerate}

The trade rule is more complicated and will not be fully described here, but see the footnote above for a concise explanation, and \cite{epsteinGrowingArtificialSocities1996}, p. 105 for a detailed one: essentially, when the MRS crossing condition is met, the price is calculated as the geometric mean of the two agents' MRS's, and quantities to be exchanged are determined by those prices.  Trade then occurs as long as the MRS crossing condition is not violated and welfare increases for both agents.

\begin{center}
\begin{figure}[h!]
\centering
\includegraphics[scale=0.8]{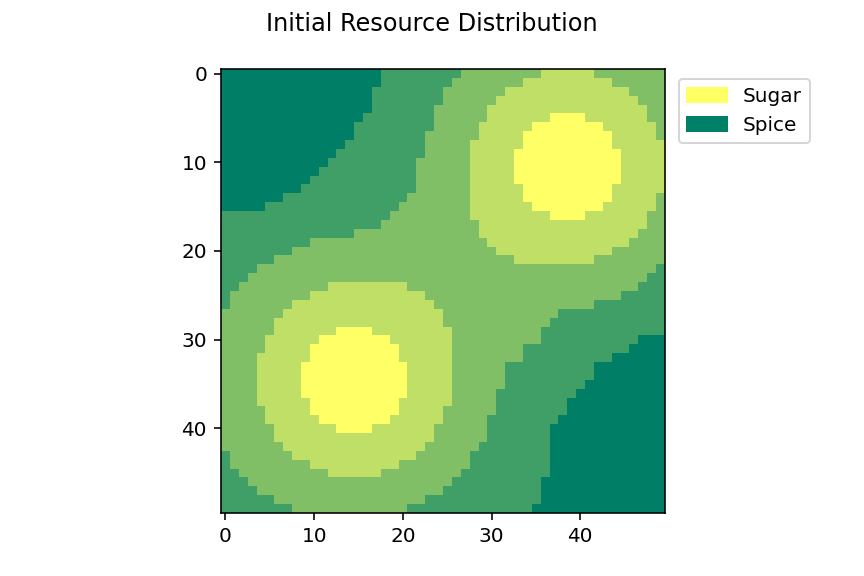}
\caption{Distribution of pollution, $t$ = 0}
\label{sugar_dist}
\end{figure}
\end{center}

Figure \ref{sugar_dist} displays the initial distribution of sugar and spice in the model.  As in the original formulation by Epstein and Axtell, there are two sugar ``hills'' and two spice ``hills,'' one in each quadrant of the grid.

We now add an additional principle to the model: sugar is a dirty good, in the sense that both the harvesting and the consumption of sugar produces pollution on the grid cell where this occurs.  In this sense, sugar has both a production and consumption externality associated with it.  Spice does not produce any pollution.  Again, as in Epstein and Axtell's initial formulation, the pollution rule is simple: pollution occurs at rate $\beta$.  If $\rho$ is the metabolic rate of sugar for a given agent, and $s$ is the amount of sugar harvested on a grid cell,\footnote{Only one unit of sugar is consumed at each time step of the model, so this quantity drops out of the equation.} then total pollution at time step $t$, $p_t$, is given by:

\begin{equation}
p_t=p_{t-1}+\beta(s+\rho).\label{eq_pol_dynamics}
\end{equation}

\begin{center}
\begin{figure}[h!]
\centering
\includegraphics[scale=0.8]{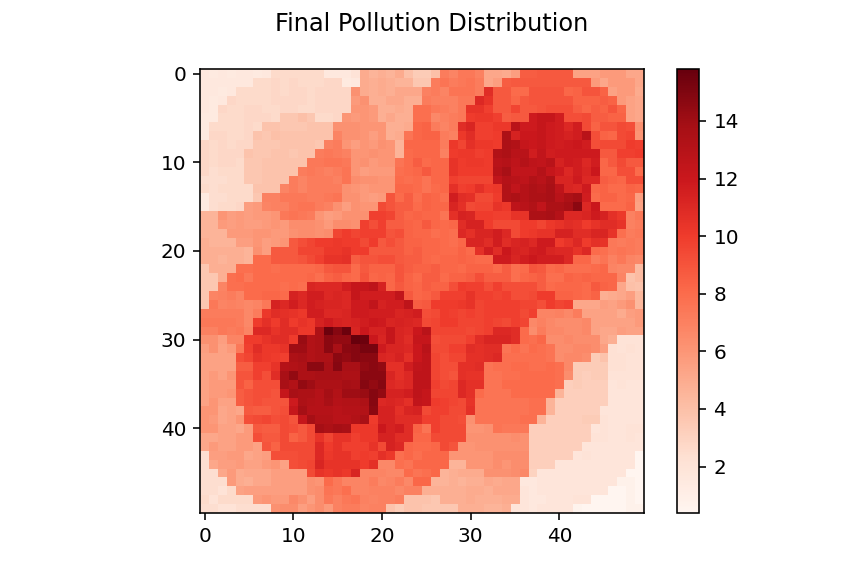}
\caption{Distribution of pollution, $t$ = 500}
\label{poll_dist}
\end{figure}
\end{center}

Figure \ref{poll_dist} shows the final pollution distribution over the model lattice for one 500-step simulation.  The plot also provides a reference for the (unitless) pollution distribution, which is important when setting the bounds for possible values of our sugar production cap (see below).

Pollution has two direct effects with which we are concerned.  One, pollution adversely impacts the welfare of agents, i.e. ${\partial U}/{\partial p}<0$.\footnote{This does \textit{not} appear to be the case in Epstein and Axtell's original model.  In their formulation, pollution seems only to enter the utility function insofar as the utility function is the metric based on which decisions about where to move are made; that said, Epstein and Axtell's model did not, at baseline, return the utility of each agent as a principal measure.  They were more concerned with price dynamics.}  Second, through its effect on utility calculations, pollution impacts agents' decisions about where to move, as cells with more pollution are less desirable.  In particular, as in the utility function in general, it is the ratio of sugar (and spice) to pollution which agents care about.

It is worth noting that, despite the fact that the utility function for agents can be written down, in general a closed form of the model does not exist.  This is in part due to the randomness inherent in the initial configuration of the model, which makes it difficult to predict \textit{a priori} how emergent properties of the system, like the welfare distribution, will evolve over time.  At the same time, agents have imperfect information because they cannot ``see'' the state of the entire lattice at any given point of time, and are in fact heterogeneous over how imperfect this information is (specifically, how far out their vision permits them to see) and over other characteristics, such as their metabolic requirements for survival.  In this sense, we can refer to the Sugarscape model as a ``black box,'' and shall do so throughout this paper: although a set of rules can be written down that may help interpret why certain results and features of the model emerge, neither total prediction nor total understanding of causal mechanisms is generally possible with this complex version of the model.

\subsection{Policy Options}\label{policy_options}

\begin{center}
\begin{figure}
\centering
\includegraphics[scale=0.7]{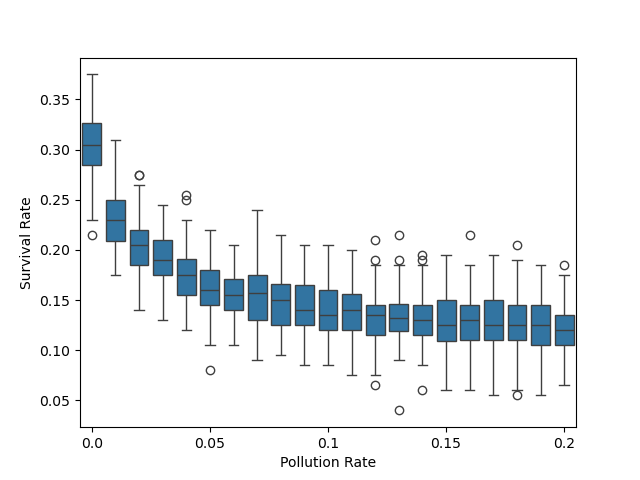}
\caption{Each boxplot is the distribution of agent survival rates across 100 model runs, of 500 steps each, of the sugarscape trade model.}
\label{poll_surv}
\end{figure}
\end{center}

\begin{center}
\begin{figure}
\centering
\includegraphics[scale=0.7]{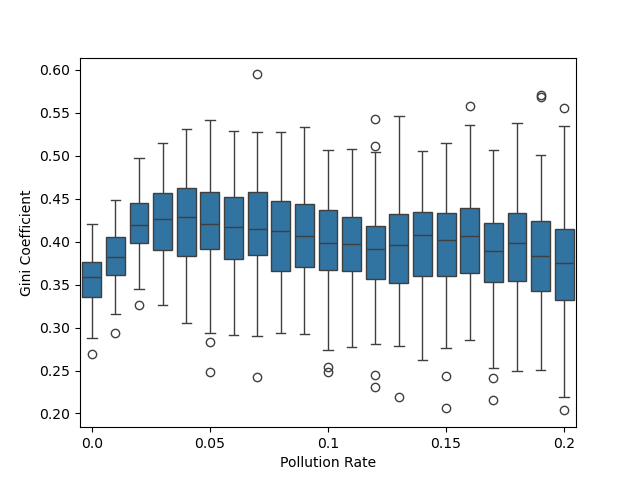}
\caption{Each boxplot is the distribution of Gini coefficients (calculated from agent welfare) across 100 model runs, of 500 steps each, of the sugarscape trade model.}
\label{poll_gini}
\end{figure}
\end{center}

As one may expect from both real experience and prior economic theory, the introduction of pollution qualitatively changes model outputs and processes, in general to the detriment of the agents in the model.  Figure \ref{poll_surv}, for example, shows that as the pollution rate increases, fewer agents tend to survive to the end of the model run.  Figure \ref{poll_gini} demonstrates that, furthermore, the Gini coefficient, a measure of inequality that is increasing in the skewness of the welfare distribution in the agent population (see, e.g., \citealp{lermanNoteCalculationInterpretation1984}), can be quite high at increased pollution rates.

We thus introduce several policy parameters into the model which aim, to one degree and in one way or another, to mitigate the levels of pollution in the lattice and/or pollution's impact on the agents.  The four policy levers are:

\begin{enumerate}
\item \textbf{Production cap on sugar}: A simple cap on the level of sugar that can be harvested by an agent in a single period.  The cap is based on the level of pollution in a given cell: if the pollution level is above the given threshold, no sugar may be harvested on the cell in that period.
\item \textbf{Reinvestment subsidy to spice}: Some amount of the capped sugar growth is assumed to be reinvested as a small subsidy of spice to the agent; that is, when this parameter is nonzero, capped sugar growth is viewed as growth that can be diverted back to the ``consumer'' (i.e., an agent).  This occurs at a constant level--if no sugar can be harvested because pollution is too high in a cell, a specified amount of spice is given back to the consumer.\footnote{Unless otherwise stated, for all models run below, this policy was not optimized but fixed at a level of 0.5.  We list it here for completeness but thus consider only the other three main policy options for the reasons described in the next section.}
\item \textbf{Tax on the trade of sugar}: A simple tax on the price of sugar for each sugar-spice trade that occurs.
\item \textbf{Tax on sugar consumption}: An agent must pay extra to consume sugar (based on their metabolic rate), however this amount is returned to them as extra spice; this is thus another form of ``limit-and-reinvest.''
\end{enumerate}

The policy options were chosen to reflect both a variety of real-life environmental policies as well as differing levels of intervention with differing degrees of effects on agent welfare and survival.  For example, policy option (1) should have quite a large impact relative to the others with regards to agent survival, because it directly limits the amount of one of the goods that can be consumed by the agent.  On the other hand, it finds a real-life counterpart in, for instance, hard limits on carbon emissions that are favored by many experts for keeping anthropogenic global warming below a certain level.  A tax on sugar trade would be similar in this sense to a tax on the actual exchange of carbon-intensive goods, while the tax on sugar consumption would be analogous to a tax that targets consumers of such goods but returns this tax to them as a subsidy for clean alternatives (e.g., imagine a tax on disposable products that funds a subsidy for reusable alternatives).

\section{Results}\label{results}

We now apply the proposed statistical framework for sensitivity analysis and policy optimization to the above Sugarscape model, to investigate optimal combinations of policy options described in subsection \ref{policy_options}. In what follows, we employ the optimization of several different objective metrics, including mean welfare in the population of agents, survival rate of the agents, and the welfare distribution in the population of agents. We first consider sensitivity testing results, then investigate how such sensitivity affects optimal policy decisions and model dynamics.

\subsection{Sensitivity Testing}\label{sens_results}

We first test, using the procedure outlined in subsection \ref{sensitivity_analysis}, whether the considered policy options are indeed sensitive to each of the three state variables: the pollution rate, the minimum initial endowment level of an agent (i.e., the lowest possible amount of resources that an agent can randomly start with), and the maximum metabolism rate of an agent (i.e., the highest level of consumption required for an agent to survive, which is also randomly assigned). The pollution rate was chosen here because, naturally, we may expect that how much pollution is generated by each unit of production and consumption will impact how much ``policy forcing'' is required to achieve a given level of the objective.  Similarly, if agents have to consume more or start with a lower wealth, a given objective may be harder to reach with the same policy settings. This sensitivity is investigated for each of the three considered objectives: namely, survival rate, welfare, and the Gini coefficient. To briefly summarize our results, we find robustly that all objectives are sensitive to all three of the state variables, with the exception of two cases, both relating to the minimum initial endowment.

Tables \ref{pollrate_sens}-\ref{metmax_sens} show, for each of the three state variables, the corresponding sensitivity test statistics $\Lambda$ and their p-values for each choice of objective, i.e., survival rate (SR), welfare (W), and Gini coefficient (G). This sensitivity test was performed using $n=40$ evaluation points, following the rule-of-thumb in section \ref{sec:algorithmic_statement}. The test p-value is evaluated from a chi-squared distribution with $\nu = d + p = 4$ degrees-of-freedom.  The simulation of such data from the ABM required just under 11.5 hours of computation time (parallelization was not possible here due to the nature of the experiments run, but 40 cores and 120 GB of memory were used nevertheless).  This shows that, even for a relatively simple ABM, the computational cost of data generation can be quite high.

For two of the state variables, pollution rate and maximum metabolism, we see that the test p-values are all nearly zero (and certainly below the significance level of $\alpha = 0.05$), which suggests strong statistical evidence that optimal policy choice is sensitive to both pollution rate and maximum metabolism for all objective choices. In other words, both the maximum metabolic rate and the pollution rate of the agents in the population could be considered strong moderators of the optimal policy choice, when considering survival rate, welfare, and inequality as objectives to optimize.

The case for the minimum agent endowment (see Table \ref{endowmin_sens}) is more nuanced. From this table, we see that when welfare is the objective, the test p-value is nearly zero, which suggests that the optimal policy can be highly sensitive to the minimum endowment. This makes sense, since a society with a different distribution of minimum wealth levels but the same distribution of maximum wealth levels will naturally have different average welfare levels (which is our main welfare measure). In other words, if there are more or less overall resources in the population at the start of the model run, we should expect this to change overall welfare by the end of the model run, which is where we measure sensitivity. When inequality or survival rate is the objective, however, we see that the test p-values can be large (above the significance level of $\alpha = 0.05$), which suggests a lack of statistical evidence for sensitivity. One reason is that, as will be demonstrated in our dynamic analysis in subsection \ref{dyn_results}, inequality and survival rate very quickly settle into an equilibrium during any given model run.  This means that shifts in the bottom of the starting wealth distribution should matter less for the end states of these variables, as seen in the fact that for survival rate and the Gini coefficient, the minimum endowment is not a strong moderator of optimal policy.  Unlike the pollution rate and metabolism, the endowment minimum only determines the starting amount of resources, but does not impact agent behavior \textit{at each model step}.  Thus, the result that model end states are not as strongly moderated by the endowment variable is quite intuitive.

The above analysis shows that the optimal policy for the considered model is indeed sensitive to various state variables (though for minimum endowment, less sensitive for certain objectives), if our interest lies in the end state of the model. Investigating how this optimal policy changes over the state variable space is thus of interest. This is particularly important when there are uncertainties associated with the considered state variables. For example, when empirically calibrating the pollution rate from real-world data, one may have considerable uncertainty on the estimated parameter, and exploring how the optimal policy changes for different plausible choices of this rate is thus important. An efficient method for such exploration is our proposed Bayesian optimization approach.

\subsection{Policy Optimization}\label{bo_results}

In the following, we consider jointly optimizing over three of the policy variables described above: the trade tax, the consumption tax, and the production cap on sugar. Using the notation from section \ref{bo}, these policy variables will be referred to as $x$.  The reinvestment amount is fixed at 0.5 to simplify analysis, since it would of course be optimal to convert as much sugar as possible into spice and return it to the consumer (i.e., the policy should not be binding).  However, if this conversion is thought of as a technology that is unlikely to change over the course of a model run, and if the regrowth rate of resources is constant across cells and time, then it makes sense to leave this policy at a constant value that is not dependent on the cell's total amount of each resource nor the agents' metabolism.  Furthermore, to improve numerical stability, the range of possible values for the taxes was narrowed down to the closed interval between 0 and 1.  For the sugar production cap, we test values between 6 and 15.\footnote{These bounds were chosen heuristically based on results from prior model runs; for instance, it was observed that setting the cap much below 6 resulted in all agents dying in most model runs.}  Note that a higher cap reflects a \textit{less} binding policy, as this allows for more pollution on each grid cell. The pollution rate is fixed at 20\% for all model runs.

\begin{center}
\begin{figure}
\centering
\includegraphics[scale=0.7]{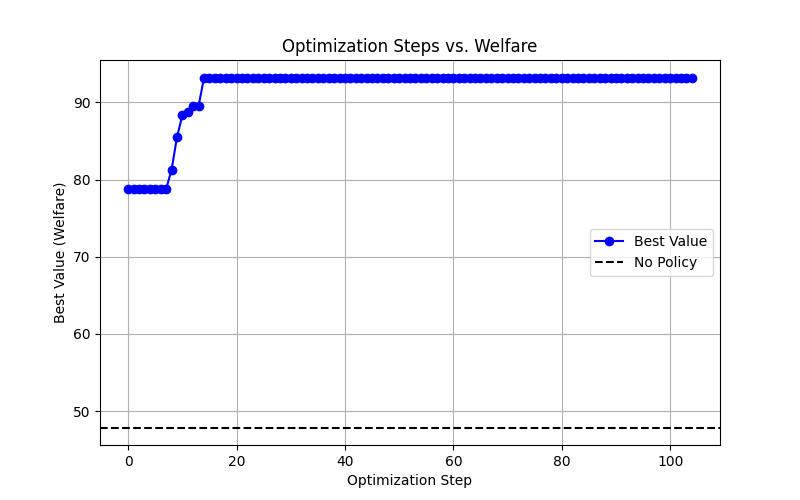}
\caption{A plot of the best observed welfare objective using the proposed Bayesian optimization procedure as a function of the number of sequential iterations. The welfare objective for the ``no policy'' scenario is shown by the dotted line.}
\label{welfare_itr_plot}
\end{figure}
\end{center}

\begin{center}
\begin{figure}
\centering
\includegraphics[scale=0.7]{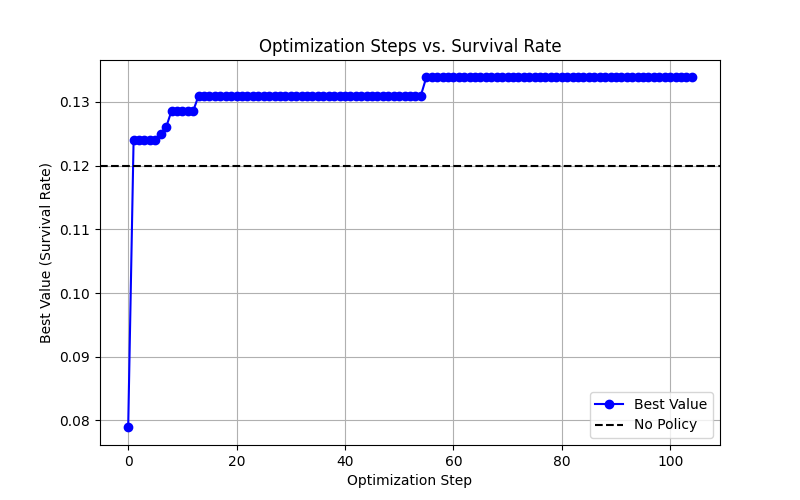}
\caption{A plot of the best observed survival rate objective using the proposed Bayesian optimization procedure as a function of the number of sequential iterations. The survival rate for the ``no policy'' scenario is shown by the dotted line.}
\label{survival_rate_itr_plot}
\end{figure}
\end{center}

\begin{center}
\begin{figure}
\centering
\includegraphics[scale=0.7]{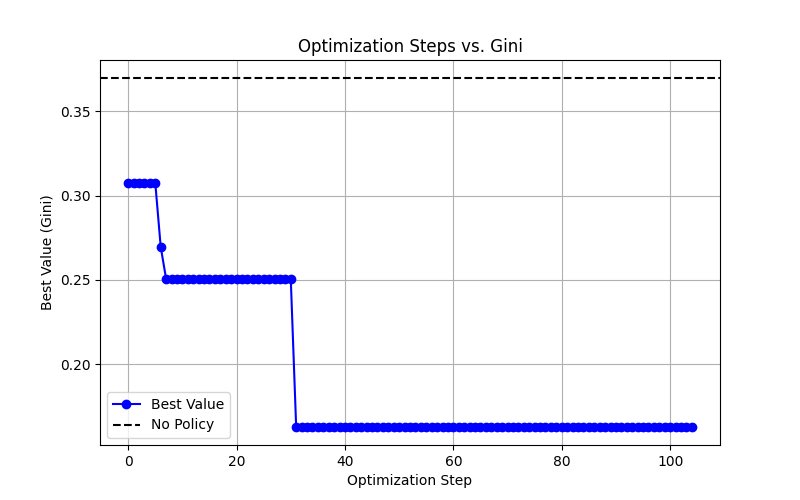}
\caption{A plot of the best observed Gini coefficient objective using the proposed Bayesian optimization procedure as a function of the number of sequential iterations. The Gini coefficient for the ``no policy'' scenario is shown by the dotted line.}
\label{gini_itr_plot}
\end{figure}
\end{center}

Figures \ref{welfare_itr_plot}-\ref{gini_itr_plot} show the progression of our Bayesian optimization algorithm (Algorithm \ref{alg:bo}), using $m=5$ initial evaluation points and $M = 100$ subsequent sequential evaluation points. As an initial pass, three different objectives (unconstrained) were investigated.  The first is a simple welfare maximization procedure, where social welfare is taken as the average welfare across all agents.  The second is maximization over the survival rate of agents in the model, calculated simply as the number of agents still alive at the end of the 500-step model run divided by the initial number of agents (200 in all cases).  Finally, we compute and minimize the Gini coefficient of our simulated society using the welfare distribution in the surviving agent population, i.e., we aim to minimize the inequality in the population. Each figure shows the optimization of a different objective. In these plots, the black horizontal dashed lines show the computed value of the objective when \textit{no} policy is implemented; that is, when both taxes are set to 0 and the production cap is set to infinity.

These plots reveal several interesting observations. First, in all cases, we see that the Bayesian optimization procedure converges quite quickly to an optimal solution; such a solution is generally found within 60 sequential iterations of the procedure, with little-to-no improvements made in subsequent iterations. This is typical of BO algorithms, which can find optimal solutions with limited evaluations of the black-box objective function, as long as there are not many variables to optimize over (\citealt{chen2024hierarchical}). Second, the optimized policies offer considerably improved objectives over the no-policy baseline, which shows the potential of BO for identifying promising new policy options from the black-box ABM.  Interestingly, such an improvement is smallest for the survival rate.  As is noted in the following subsection, survival rate tends to be particularly low in the long-run regardless, at the very least, of trade tax level, potentially due to the weakness of some instruments relative to the objective. Finally, it is important to note here the computational cost of ABM simulation for this optimization procedure. To optimize over welfare, for example, evaluations of the black-box objective $f$ required nearly 3 hours of computation, on a 32-core processor with 64 GB of memory. This shows that, even for a relatively simple ABM, the computational cost of evaluating the desired objective can make policy optimization highly challenging; our Bayesian optimization approach provides a solution to this bottleneck. 

\begin{center}
\begin{figure}
\centering
\includegraphics[scale=0.7]{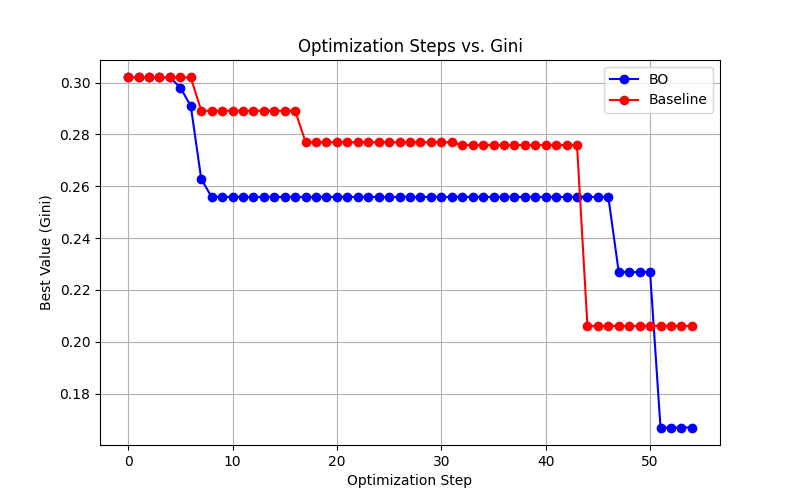}
\caption{A plot of the best observed Gini coefficient objective using the proposed Bayesian optimization procedure and the random sampling baseline, as a function of the number of sequential iterations.}
\label{gini_comp_plot}
\end{figure}
\end{center}

To gauge the effectiveness of our BO procedure, another baseline method may be to simply select policy points at random over the decision space, run the ABM at such points to evaluate the objective $f$, then take the tested policy point with the best simulated objective. Figure \ref{gini_comp_plot} shows the comparison of BO with this random benchmark for optimizing the Gini coefficient, where both methods are compared on $M=50$ sequential iterations to speed up computation. We see that, at the end of the sequential procedure, the optimal policy found by the random baseline has a considerably higher Gini coefficient (around 0.205) than that found by the BO procedure (around 0.170). Indeed, throughout the sequential procedure, BO largely dominates the random baseline; there is only a brief period where the random baseline yields better performance due to chance. This improvement of BO over random points is in line with the literature (\citealt{frazier2018tutorialbayesianoptimization}), and highlights the importance of a carefully-trained acquisition function to guide the sequential optimization of a black-box ABM.

Next, we inspect the corresponding optimal policies found by the BO algorithm for each choice of objective, which are reported in Table \ref{opt_params}. Again, several interesting observations can be gleaned. First, for all objectives, the optimized policy has very low levels of trade tax.  This suggests, particularly in light of the findings regarding this tax in the next section, that the trade tax is a relatively weak instrument for controlling the impacts of overconsumption and pollution. One reason may be that the trade tax does not include the ``return to consumer'' scheme which the consumption tax does: the trade tax purely increases the price of doing business, without giving any windfall back to the consumer as the clean good.

Second, the optimization of survival rate does not appear to require strong policy instruments, with the cap set about midway within its range, no consumption tax, and a very low trade tax of 10\%.  However, this only leads to a final survival rate of 13\%.  On the other hand, when optimizing over the Gini coefficient, we find that the cap is binding (i.e., set to its lowest possible level of 6) and the consumption tax increases to 30\%, while the optimal trade tax only marginally decreases to 0.  The survival rate in this case, however, is a disheartening 1\%. This highlights an important finding of our analysis: there are strong tradeoffs between the three considered objectives. By targeting the optimization of the Gini coefficient, such a coefficient increases by a factor of about 2.5 (from 0.39 to 0.16), but results in a decrease in the survival rate by a factor of 13 (from 0.13 to 0.01). Similarly, by targeting the optimization of welfare, the welfare objective almost doubles (from 49.10 to 93.22), but the survival rate accordingly almost halves (from 0.13 to 0.08).  This suggests that those wishing to make optimal policy decisions must either stake their claim on a particular objective, or employ more sophisticated multi-objective Bayesian optimization techniques to achieve a desired balance between competing objectives; we discuss more on the latter point in the conclusion.

\subsection{Dynamic Analysis}\label{dyn_results}

Finally, we investigate next the dynamics of the considered ABM, namely how various states of the model change over the period of the model run. The reasons are two-fold: such an analysis provides further insights on how mitigation policies operate, and demonstrates the practical usefulness of an ABM in analyzing policy dynamics. Here, for easier analysis and interpretation, we choose to optimize only one of the three policy options at a time, although we have shown above that optimizing over multiple policy parameters is indeed possible.  Specifically, in our analysis of model transition dynamics, we optimize the trade tax on sugar, the dirty good, with respect to the \textit{overall} (i.e., at the final time step of the model relative to the start) survival rate of agents in the Sugarscape model.

\begin{center}
\begin{figure}[h!]
\centering
\includegraphics[scale=0.8]{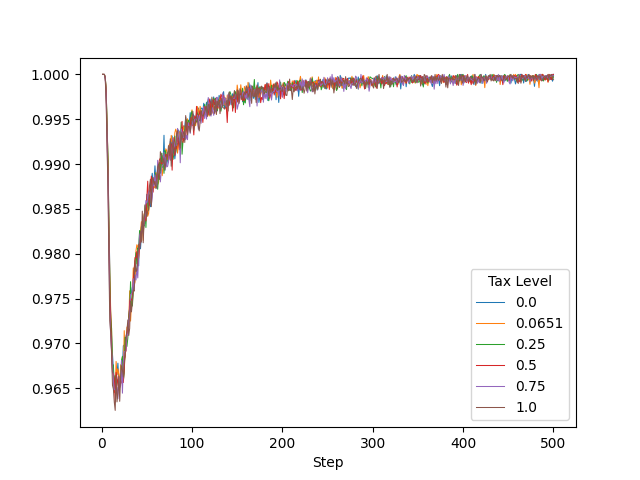}
\caption{Survival Rate Dynamics}
\label{survival_rate_dyn_full}
\end{figure}
\end{center}

The trade tax which maximizes the survival rate of agents in the model, according to our BO algorithm, is found to be approximately 6\%, a fairly nominal amount.\footnote{BO was done over $M=100$ optimization steps, using 150 model iterations at each step and averaging the objective across iterations.  As can be seen in the dynamic plots, the model itself was run for 500 time steps at each iteration.}  Figure \ref{survival_rate_dyn_full} shows the instantaneous survival rate (calculated from one period to the prior) in an ensemble of 250 model runs over time, each using a different choice of the trade tax, including the above optimized level. For all choices of tax levels, this figure shows that initially, agents begin to die off at a fairly rapid rate in all cases, before the instantaneous survival rate increases again and settles into a steady state close to 1.  Despite this potentially optimistic portrait, however, we find that throughout most model runs which include the polluting good, most agents will die by the end of 500 periods.  Even at the optimal trade tax, the overall survival rate found during our optimization procedure is less than 14\%. As has been discussed above, the survival rate is only weakly improved by our Bayesian optimization algorithm; although we find that uncertainty over several state variables (i.e.  pollution rate and agent metabolism, as discussed in subsection \ref{sens_results}) strongly \textit{moderates} the relationship between policy parameters and survival rate, the weakness of said instruments (in particular, the trade tax) may mean that direct changes to them lead to only miniscule differences in survival rate.

\begin{center}
\begin{figure}[h!]
\centering
\includegraphics[scale=0.8]{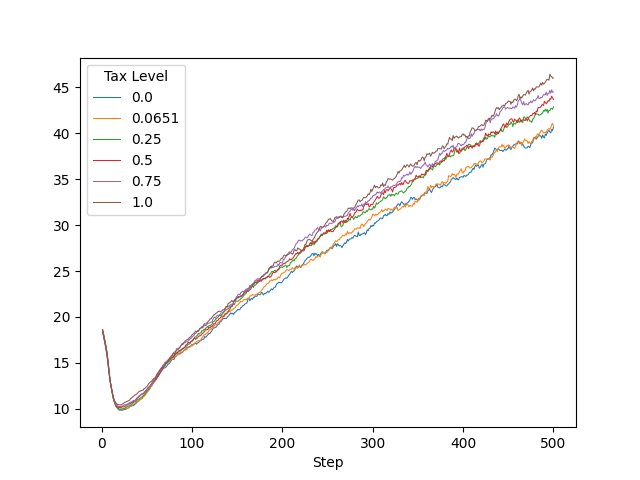}
\caption{Median Welfare Dynamics}
\label{med_welfare_dyn_full}
\end{figure}
\end{center}

\begin{center}
\begin{figure}[h!]
\centering
\includegraphics[scale=0.8]{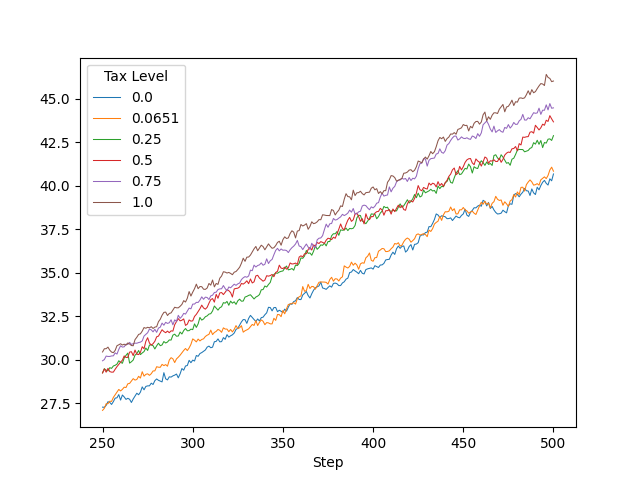}
\caption{Median Welfare Dynamics}
\label{med_welfare_dyn_end}
\end{figure}
\end{center}

There also appears to be a tradeoff between survival rate and welfare objectives in the population of agents.  If we perform the Bayesian optimization procedure with welfare as the objective, the optimized policy leads to a survival rate closer to 11\%, lower than the 14\% reported above; however, average welfare in the population in that case increases from about 48 to about 56. Figure \ref{med_welfare_dyn_full} shows the path of the median welfare in the (surviving) population of agents at each time step of the model, and figure \ref{med_welfare_dyn_end} hones in on the final 250 time periods.  Unlike survival rate, welfare appears to monotonically increase over time and never settles into a steady state.  Additionally, the median welfare is roughly increasing in the trade tax, with the tax level optimized to the survival rate further having a relatively low transition path for welfare.  While this result may at first seem confusing, given that traditional economic theory holds that taxes will be distortionary and thus more likely to decrease welfare, the specific setup of this model may explain this finding.  Since the resources in the model are essentially common pool resources, in the sense that there is no limit on access to them, an additional distortion is at play which the tax may partially correct.  This explanation is in line with the intercountry trade model of \cite{branderOpenAccessRenewable1998}, which finds that with a common pool resource being exchanged, traditional ``gains from trade'' are lost due to the overexploitation of the resource, such that limiting trade can in fact be welfare-improving.  In this situation, there may be further externality arising from the polluting nature of sugar, such that the tax is mitigating the welfare-diminishing effects of pollution.

\begin{center}
\begin{figure}[h!]
\centering
\includegraphics[scale=0.8]{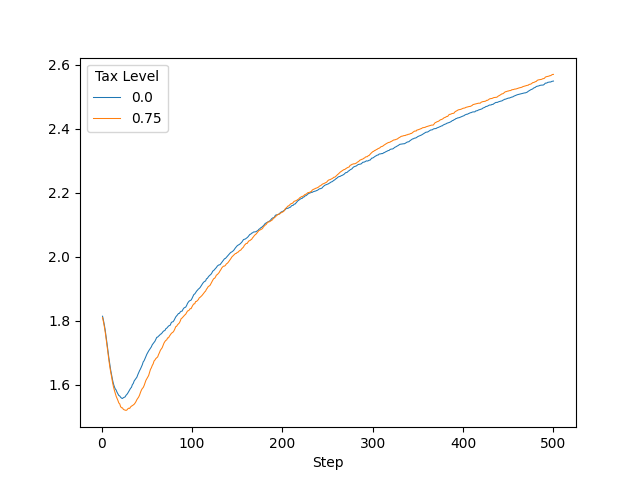}
\caption{Decomposition of Sugar Dynamics}
\label{sugar_decomp_dynamics}
\end{figure}
\end{center}

\begin{center}
\begin{figure}[h!]
\centering
\includegraphics[scale=0.8]{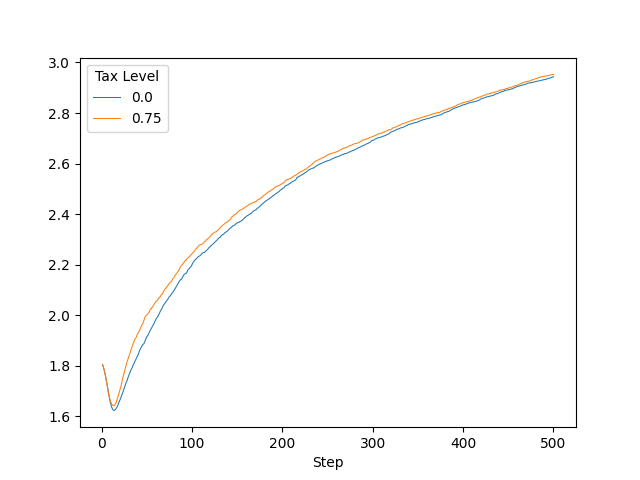}
\caption{Decomposition of Spice Dynamics}
\label{spice_decomp_dynamics}
\end{figure}
\end{center}

To further investigate how the trade tax operates, we decompose its effects on welfare into two components: a consumption effect from sugar, and another from spice.  Since the utility function used is Cobb-Douglas (\citealt{brownCobbDouglasFunctions2018}) in form, we can do this by simply taking the logarithm of utility (disregarding for now the negative component from pollution) and decomposing this into two additive effects, one from each of the two goods.  Figures \ref{sugar_decomp_dynamics} and \ref{spice_decomp_dynamics} show how welfare from the consumption of the two goods changes over the course of an ensemble of 150 model runs, taking the average over agents and model runs.  Here, lines in the plot represent a case where the trade tax is set to 0 (a no-policy counterfactual scenario) and another case in which it is set to a moderately high level of 75\%.  Interestingly, for the dirty good (sugar), the tax level initially decreases gains from consumption, which is expected since trade of that good is being taxed. Around time step 200, this pattern reverses: the tax increases the amount of welfare that agents are receiving from the dirty good.  This rather counterintuitive result can also be partly explained by the economic theory of open access natural resources.  As \cite{branderInternationalTradeOpenAccess1997} discuss in the case of a simple two-country model, liberalized trade regimes (which correspond to the no-tax scenario here) can initially lead to welfare gains, but often result in long-term welfare losses as the open-access resource is depleted at a higher rate. This indeed seems to be the case for our taxed, dirty good (sugar).  For the clean good, welfare gains are consistently lower for the no-tax scenario, ostensibly because of substitution effects: the trade tax leads consumers to increase their consumption of the clean good, as a policymaker would desire and as one might expect.  Thus, perhaps unsurprisingly, we find that we can reproduce some basic economic results using even this simple model and one of our policy instruments.  Furthermore, we have shown that the trade tax does seem to lead to substitution away from the dirty good and, in the long-run, welfare gains as the open access problem is partially ``solved.''

\begin{center}
\begin{figure}[h!]
\centering
\includegraphics[scale=0.8]{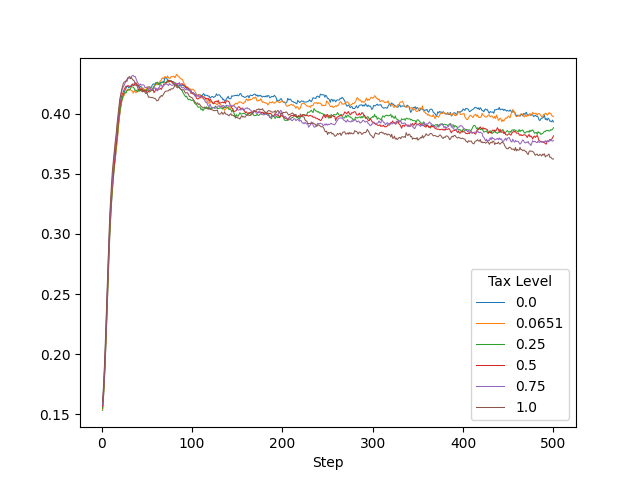}
\caption{Gini Coefficient Dynamics}
\label{gini_dyn_full}
\end{figure}
\end{center}

Another notable result is that for the survival rate, steady state is approached very quickly.  On the other hand, for the Gini coefficient, as shown in figure \ref{gini_dyn_full}, steady state is never truly reached: the Gini coefficient increases very rapidly at first, then begins to decrease slowly.  It could be argued that the dynamics are so slow after the first few time periods that the welfare distribution has essentially reached an equilibrium point, but it is noteworthy that as agents continue to die off, the resulting smaller society becomes more and more egalitarian.  This rather troubling result should not, however, be taken at face value, both because the decreases in the Gini coefficient are small and because in this heavily stylized model, it makes sense that it is easier to ``equalize wealth'' in such a small society of agents.  In the real world, there is no reason to believe \textit{a priori} that this would be the case.  Nevertheless, this again points to interesting tradeoffs between objectives that our model suggests.

\section{Conclusion}\label{conclusion}

In this paper, we have proposed a novel statistical framework for sensitivity testing and policy optimization using costly agent-based models (ABMs). While recent developments in ABMs allow one to reliably model complex intervention behavior, their use for policy optimization is hindered by the fact that ABM simulations can be computationally expensive. Our framework addresses this via the use of machine learning techniques that enable effective modeling and optimization performance with limited simulation runs. We proposed a hypothesis testing approach for investigating sensitivity of the optimal policy using flexible Gaussian process models, then outlined a Bayesian optimization reinforcement learning procedure for policy optimization. We then investigated the effectiveness of this framework using an extension of the classic Sugarscape model, where we showed that our identified policies improve upon baseline approaches (including random search approaches) for optimizing a target objective, with insightful sensitivity and dynamic analyses that relate back to economic theory.


While the policies explored here are simple, they do have parallels in policy discourse around, e.g., the limitation of greenhouse gas emissions.  In an empirically calibrated model, such as the ones described in section \ref{review}, it is reasonable to assume that policies will be more complex.  This presents several challenges to the modeler/policymaker.  First, it may become difficult, as we have done, to use simple heuristics based on a few model runs to bound policy parameters in any meaningful way.  Second, the topology of the relevant ``policy spaces'' may be unclear to begin with: for example, should a policy meant to subsidize fertilizer purchases for farmers be constrained by budgetary considerations?  Should it place requirements on the type of fertilizer purchased in order to meet sustainability demands?  Finally, computational challenges are likely to be more pressing with an empirically calibrated model.  More agents, more parameters and variables, and more complex behavioral rules all mean greater computation time.

The ability of our Bayesian optimization approach to quickly identify good policy choices with limited simulation runs shows promise for the optimization of ABMs for broad applications. We hope that our methodological contributions here will inspire researchers in the policy space to more seriously consider using agent-based models, especially where heterogeneity, dynamics, and spatial effects are thought to be important.  In future work, we hope to expand upon these methods further, by considering both the uncertainty underlying state parameter estimates and the possibility of multiple optima with diverse black-box optimization techniques (\citealt{miller2024expected}).  We will further explore the use of multi-objective Bayesian optimization techniques (\citealt{daulton2022multiobjectivebayesianoptimizationhighdimensional}), which facilitates the optimization of potentially competing objectives from the black-box ABM. The exploration of deeper variants of GPs (e.g., \citealt{li2025additive}) can also accelerate the identification of optimal policies in high-dimensional decision spaces. Finally, it would be worthwhile to add constraints to our objectives, e.g., the requirement that total pollution be kept below a specified level, to better reflect realistic policy optimization problems. Such directions will further the applicability of sophisticated ABMs as a robust tool for designing targeted and effective environmental policies.

\section*{Acknowledgments}

The authors would like to thank Derek Cho, Mark Borsuk, Marc Jeuland, Martin Smith, James Moody, and Robyn Meeks for excellent constructive feedback and suggestions.

\section*{Declaration of Competing Interests}

The authors declare no potential conflicts of interest with respect to the research, authorship and/or publication of this article.

\bibliographystyle{chicago}
\nocite{*}\bibliography{ems_bib}

\clearpage

\section*{Tables and Figures}

\begin{center}
\begin{figure}[h!]
\centering
\tikzstyle{box} = [rectangle, rounded corners, minimum width=3cm, minimum height=1cm,text centered, draw=black]
\tikzstyle{box_abm} = [rectangle, rounded corners, minimum width=3cm, minimum height=1cm,text centered, draw=black, fill = green]
\tikzstyle{arrow} = [thick,->,>=stealth]
\begin{tikzpicture}[node distance=2cm]
\node (theory) [box, text width = 2cm] {Political and economic theory};
\node (discuss1) [box, right of = theory, xshift = 2cm, text width = 2cm] {Stakeholder discussions};
\node (evidence) [box, right of = discuss1, xshift = 2cm, text width = 2cm] {Empirical evidence};
\node (bounds) [box, below of = discuss1, yshift = -2cm, text width = 4cm] {Bounds and calibration};
\node (abm) [box_abm, below of = bounds, yshift = -1.5cm, text width = 4cm] {ABM and policy optimization};
\node (discuss2) [box, below of = abm, yshift = -1.5cm, text width = 4cm] {Stakeholder discussions};
\node (test) [box, below of = discuss2, yshift = -1.5cm, text width = 4cm] {Testing/trial of policy};
\node (implementation) [box, below of = test, yshift = -1.5cm, text width = 4cm] {Implementation of policy};
\draw [arrow] (theory) -- (bounds);
\draw [arrow] (discuss1) -- (bounds);
\draw [arrow] (evidence) -- (bounds);
\draw [arrow] (bounds) -- (abm);
\draw [arrow] (abm) -- (discuss2);
\draw [arrow] (discuss2) -- (test);
\draw [arrow] (test) -- (implementation);
\draw [arrow] (implementation) -| (theory);
\draw [arrow] (implementation) -| (evidence);
\end{tikzpicture}
\caption{A basic framework for understanding the potential role of ABMs (and similar models) in policymaking.  The arrows on the side of the figure emphasize that this is an iterative process in which policies are continually redesigned as new information about their effects is revealed and modeling techniques/computational power improves.}
\label{schematic}
\end{figure}
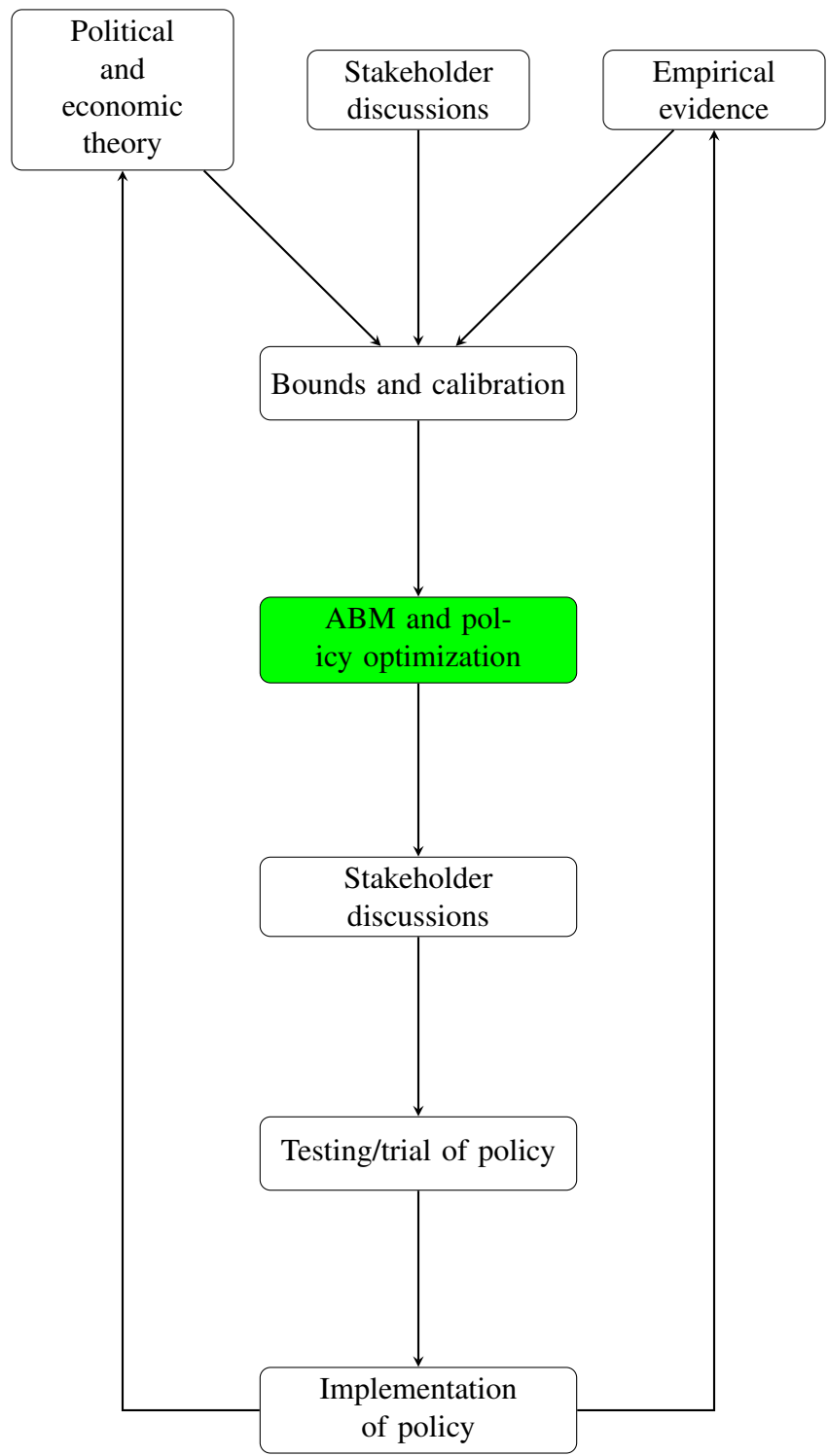
\end{center}

\begin{table}[ht]
\centering
\caption{The test statistics $\Lambda$ and corresponding p-values of our sensitivity test for the three considered objectives, when the state variable is taken as pollution rate.} \vspace{-0.4cm}
\begin{tabular}{| c | c | c |}
\hline
$y$ & $\Lambda $ & p-val \\
\hline
SR & 185.96 & .000 \\
W & 2136 & .000 \\
G & 90.38 & .000 \\
\hline
\end{tabular}
\label{pollrate_sens}
\end{table}

\begin{table}[ht]
\centering
\caption{The test statistics $\Lambda$ and corresponding p-values of our sensitivity test for the three considered objectives, when the state variable is taken as endowment minimum.} \vspace{-0.4cm}
\begin{tabular}{| c | c | c |}
\hline
$y$ & $\Lambda $ & p-val \\
\hline
SR & 7.1 & \textbf{.131	} \\
W & 1718 & .000 \\
G & 0.6 & \textbf{.963} \\
\hline
\end{tabular}
\label{endowmin_sens}
\end{table}

\begin{table}[ht]
\centering
\caption{The test statistics $\Lambda$ and corresponding p-values of our sensitivity test for the three considered objectives, when the state variable is taken as maximum metabolism.} \vspace{-0.4cm}
\begin{tabular}{| c | c | c |}
\hline
$y$ & $\Lambda $ & p-val \\
\hline
SR & 746.46 & .000 \\
W & 1236 & .000 \\
G & 106.78 & .000 \\
\hline
\end{tabular}
\label{metmax_sens}
\end{table}

\FloatBarrier

\FloatBarrier

\begin{table}[h!]
\centering
\caption{Optimal policies identified by the proposed Bayesian optimization procedure.}
\begin{tabular}{| c | c | c | c | c | c | c |}
\hline
\textbf{Objective} & \textbf{Trade Tax} & \textbf{Cap} & \textbf{Consumption Tax} & \textbf{Welfare} & \textbf{Survival Rate} & \textbf{Gini} \\
\hline
Welfare & 0 & 9.13 & 0.83 & 93.22 & 0.08 & 0.32 \\
Survival Rate & 0.1 & 13.27 & 0 & 49.1 & 0.13 & 0.39  \\
Gini & 0 & 6 & 0.3 & 31.53 & 0.01 & 0.16 \\
\hline
\end{tabular} 
\label{opt_params}
\end{table}

\appendix

\setcounter{table}{0} \renewcommand{\thetable}{A.\arabic{table}}
\setcounter{figure}{0} \renewcommand{\thefigure}{A.\arabic{figure}}
\setcounter{section}{0} \renewcommand{\thesection}{A\arabic{section}}
\setcounter{equation}{0}

\clearpage

\end{document}